\documentclass[twocolumn, 10pt, amsfonts, amsmath, amssymb, superscriptaddress, noeprint, aps]{revtex4-2}

\usepackage{pdfpages}
\makeatletter
\AtBeginDocument{\let\LS@rot\@undefined}
\makeatother

\usepackage{microtype}
\usepackage{lmodern}
\usepackage[hidelinks]{hyperref}
\usepackage{graphicx}
\usepackage{placeins}
\usepackage{booktabs}
\usepackage{chemmacros}
\usepackage{siunitx}
\usepackage{physics}
\usepackage{soul}
\usepackage{xcolor}

\begin{document}
\author{Waldemar Svejstrup}
\affiliation{Center for Quantum Devices, Niels Bohr Institute, University of Copenhagen, Copenhagen, Denmark}

\author{Andrea Maiani}
\affiliation{Center for Quantum Devices, Niels Bohr Institute, University of Copenhagen, Copenhagen, Denmark}

\author{Kevin Van Hoogdalem}
\affiliation{Microsoft Quantum Lab Delft, Delft University of Technology, Delft, Netherlands}

\author{Karsten Flensberg}
\affiliation{Center for Quantum Devices, Niels Bohr Institute, University of Copenhagen, Copenhagen, Denmark}

\title{Orbital-free approach for large-scale electrostatic simulations of quantum nanoelectronics devices}

\date{February 2023}

\begin{abstract}
% 1 General field
The route to reliable quantum nanoelectronic devices hinges on precise control of the electrostatic environment.
%2 Specific field
For this reason, accurate methods for electrostatic simulations are essential in the design process. The most widespread methods for this purpose are the Thomas-Fermi approximation, which provides quick approximate results, and the Schrödinger-Poisson method, which better takes into account quantum mechanical effects.
%3 Problem addressed
The mentioned methods suffer from relevant shortcomings: the Thomas-Fermi method fails to take into account quantum confinement effects that are crucial in heterostructures, while the Schrödinger-Poisson method suffers severe scalability problems.
%4 Solution to the issue 
This paper outlines the application of an orbital-free approach inspired by density functional theory. By introducing gradient terms in the kinetic energy functional, our proposed method incorporates corrections to the electronic density due to quantum confinement while it preserves the scalability of a theory that can be expressed as a functional minimization problem.
%5 Impact
This method offers a new approach to addressing large-scale electrostatic simulations of quantum nanoelectronic devices.
\end{abstract}

\maketitle
\section{Introduction}
% Why OF method?
The continuous increase of complexity in quantum nanoelectronic devices has created the demand for a precise description of the electrostatic behavior of these systems, as this influences all other phenomena in the devices~\cite{Winkler2019, Mikkelsen2018, Armagnat_2019}. Solving the electrostatic problem requires the solution of Poisson's equation (PE) for the electric potential $\varphi$ coupled to a functional $n\qty[\varphi]$ that describes the electron density as a function of the electrostatic field itself
\begin{equation}
    -\div \qty( \varepsilon \grad \varphi) = \rho_\mathrm{fx} - e n\qty[\varphi]\,,    
    \label{eq:poisson_general}
\end{equation}
where $\rho_\mathrm{fx}$ is the eventual fixed charge in the device and $e$ is the electron charge.

In the context of hybrid superconductor-semiconductor quantum devices, two methods are currently widely used: the Schrödinger-Poisson (SP) and the Thomas-Fermi (TF) methods~\cite{Vuik_njp_2016, Dominguez_npjQM_2017, Woods_PRB_2018, Escribano_BJN_2018, Mikkelsen2018, Antipov2018, Armagnat_2019}. The SP method takes the quantum mechanical system into account within the Hartree approximation and provides precise results, but is computationally demanding~\cite{Armagnat_2019, Chen2010, Antipov2018}. The TF method, instead, uses the free-fermions gas model to approximate the charge distribution. In this way, it reasonably approximates the device's electrostatic configuration at a much lower computational cost. While the SP method provides more precise results, it is computationally much more demanding since it requires the diagonalization of the Hamiltonian. Memory and time requirements can be so high that it is practically impossible to use on realistic systems of relevant size~\cite{Mikkelsen2018, Bartolotti2007}. However, the TF method fails to properly consider quantum confinement effects as the energy functional is completely local. For this reason, a common practice is to first focus on the electrostatic problem and solve it with the TF method, and then use the calculated electrostatic potential to solve the quantum mechanical problem~\cite{Winkler2019, Mikkelsen2018}. We will refer to this approach, which is not strictly self-consistent, as Schrödinger-Thomas-Fermi (STF) method.

\begin{figure}
\centering
\includegraphics[width=1.0\columnwidth]{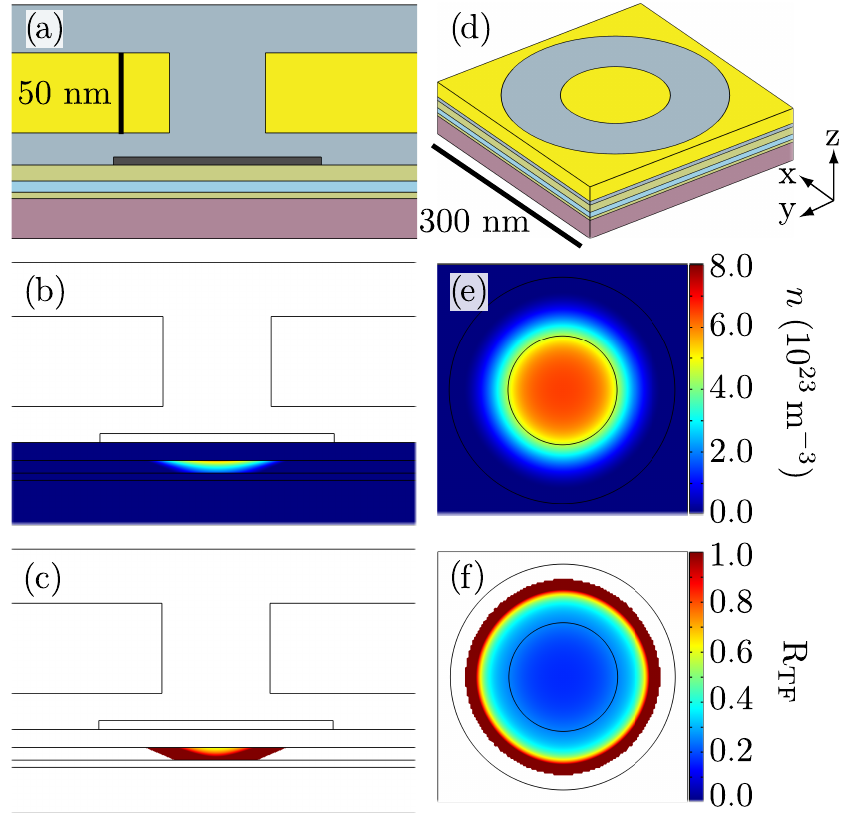}
\caption{Electrostatic simulations of two example devices. The first row shows sketches of a cross-section of a common 2DEG device (a) and a simple circular quantum dot (d). In both cases, the devices are built on top of a semiconductor stack (schematic given in Fig.~\ref{fig:lambda_calibration}) that provides the vertical confinement of electrons. The lateral confinement is controlled by \ch{Au} gates (yellow), \ch{Hf O_2} oxide (gray), and an \ch{Al} wire (dark gray). b) and e) show the electron density from a TF simulation. e) is plotted in the middle of the \ch{In As} well. c) and f) show $\mathrm{R}_\mathrm{TF}$ as defined in Eq.~\eqref{TF_error}.}
\label{fig:errors}
\end{figure}

The TF method is exact for uniform systems and holds as an approximation as long as the electron density slowly varies in space. The validity of this assumption for a specific system can be checked through the Thomas-Fermi error:
\begin{equation}
    \mathrm{R}_\mathrm{TF} \equiv \frac{\norm{\grad n (\vb{r})} }{ n(\vb{r}) k_\mathrm{F} (\vb{r}) }\,,
    \label{TF_error}
\end{equation}
where $k_\mathrm{F} (\vb{r}) = [3 \pi^2 n (\vb{r})]^{1/3}$ is the Fermi wavelength~\cite{giuliani_vignale_2005}. The TF approximation is valid in the limit $\mathrm{R}_\mathrm{TF}\ll 1$~\cite{hohenberg_kohn, Ernzerhof_JOCP_2000}. This condition is never satisfied at interfaces with vacuum or insulators where the density goes abruptly to zero. Two instances of typical devices are shown in Fig.~\ref{fig:errors}, showing the density calculated with the TF method and the value of the TF error. 

The TF method is the simplest model in the class of \emph{orbital-free} (OF) density functional theories, and some of its shortcomings are overcome in more advanced models. The defining feature of these models is that their mathematical form is exactly an energy functional of the density. This means that explicit diagonalization of the Köhn-Sham Hamiltonian is not required to determine the ground state density of the system. For this reason, OF density functional theory methods are computationally efficient and can be applied to systems with a large number of electrons. However, the accuracy of OF density functional theory is generally lower, and it is typically used for approximate calculations and to study qualitative trends. Traditional application domains of OF theories include atomic physics and material science~\cite{Dreizler1990, Engel2011, Wesolowski_2012, Karasiev_2014, Witt2018}. Similar theories are also known as \emph{semiclassical methods} in nuclear matter systems that include finite nuclei and astrophysical systems~\cite{Brack_PR_1985, semiclassical, Ring_2005, Aymard_PRC_2014}.

This paper introduces these methods in a different field: quantum nanoelectronic devices design. We adapt ideas from OF density functional theory to go beyond the TF method, improve on some of its drawbacks, and minimize any additional computational cost.
. The goal is thus to develop and assess a numerically cheap model in a variational form that includes corrections to the electronic density due to quantum confinement. Moreover, the application of these ideas to nanoelectronic devices can be far easier than in the usual application domains, like material science, since a precise estimation of excitation energies is not required. We expect such a method to better predict the charge density compared to the TF method. Estimating the electron number is especially crucial for floating metallic parts of devices where charging energy is a relevant parameter. Following the literature, we will refer to this orbital-free method as the \textit{extended Thomas Fermi} (ETF) method~\cite{semiclassical}.

A comparable theory built following similar consideration is the effective-potential potential method~\cite{Ferry_PRB_1993, Ferry_SM_2000} later expanded in the density-gradient theory~\cite{Ancona1989, Ancona2011}. Unlike our approach, the density gradient theory tries to introduce quantum-confinement corrections to the drift-diffusion equations commonly used in modeling traditional semiconductor devices operated at room temperature. Our proposal is much simpler and more tailored to quantum nanoelectronic devices operated at low-temperature with small currents in a condition very close to equilibrium.

% Applications
As a specific application domain of our method, we consider the task of optimizing gate shapes and providing an intuitive understanding of how the potential landscape is affected by variation in voltage biasing. These tasks do not require precise calculations but rather quick and responsive evaluation of a configuration. In this respect, our tool provides, in addition to a precise electric field, a model for the electron fluid which is closer to the Schrödinger-Poisson evaluations.

An inevitable feature of an orbital-free approach is the incapacity of modeling the quantization of charge. While in many applications this is a disadvantage, it can be a useful feature in certain situations. The two outputs of the method, $\varphi$ and $n$, will depend continuously on the geometrical and material parameters of the device. This means that any derived quantity $L[n, \varphi]$ will be continuous as well with respect to the parameters of the model. This allows for the possibility of developing new devices by using parametric modeling, where the function $L$ can be an objective function describing some desired property of the device (e.g., the transmission probability of a quantum point contact), allowing for computation of the optimal geometrical and material parameters by numerical optimization.

For a more precise treatment of quantization, a hybrid Schrödinger-ETF method, similar to the STF method, can be used. Additionally, the ETF method not only provides a precise estimate of the electrostatic potential $\varphi(\vb r)$ but also of the numerical density $n(\vb r)$, which can be used to generate an effective Hamiltonian through dimensionality reduction.. For example, one can be interested in projecting a 3D Hamiltonian on a 2D subspace to model a 2DEG device. This could be done by taking the weighted average of the parameters of the Hamiltonian, like, for example, the effective mass, Rashba field, electrostatic potential, or conduction band minimum.

\section{Orbital free method}
\label{sec:method}
% Method presentation
In the following, we will consider only direct bandgap semiconductors. We will neglect the hole accumulation process and model only the electron fluid. However, it is straightforward to extend the method by including a hole fluid. For low densities, we can assume that the electrons in the semiconductor are close to the $\Gamma$ point and can be reliably described by a parabolic band. In this case, a general orbital-free theory for the conduction band electrons can be expressed by an energy functional similar to the one for free electrons:
\begin{equation}
    E[n, \varphi] = K[n] + V[n] + U[\varphi] +\int_{\Omega}\dd{\vb{r}} \qty( \rho_\mathrm{fx} - e n) \varphi \,,
    \label{eq:energy_functional}
\end{equation}
where $\varphi$ is the electrostatic potential while $n$ is the numerical density of electrons~\cite{Dreizler1990, Ernzerhof_JOCP_2000, Bartolotti2007, Engel2011, VanZyl2013}. The energy functional is split into several pieces: $K[n]$ is the kinetic energy functional of the electrons, $V[n]$ is the potential energy of the electron fluid, $U[\varphi]$ is the electrostatic field energy, and finally, the last term is the coupling between the electrostatic field and the charge which comprises the free electrons charge distribution $-e n$ and the fixed charge of the system  $\rho_\mathrm{fx}$. This can be due to, e.g., doping or fixed surface charge, and can be fundamental in the electrostatic modeling of semiconductor systems~\cite{Heedt2015, Pauka2020}. The computational domain $\Omega$ can be split into three different kinds of regions: metals, insulators, and semiconductors such that  $\Omega = \Omega_\mathrm{M} \cup \Omega_\mathrm{Sm} \cup \Omega_\mathrm{I}$. 

The electrostatic field energy takes the standard form 
\begin{equation}
    U[\varphi] = \int_{\Omega} \dd{\vb{r}} \varepsilon \frac{\norm{\grad \varphi}^2}{2}\,,
\end{equation}
where $\varepsilon(\vb{r})$ is the permittivity, while the potential is 
\begin{equation}
    V[n] = \int_{\Omega_\mathrm{Sm}} \dd{\vb{r}} \qty[E_\mathrm{CBM} n + V_\mathrm{ex}(n)] \,,
\end{equation}
where $E_\mathrm{CBM}(\vb{r})$ is the conduction band minimum (CBM) of the semiconductor that acts as local chemical potential. The exchange energy can be included within the local density approximation through the  $V_\mathrm{ex}(n(\vb{r}))$ term, which we neglect in this work as including this would make the comparison with the SP method more complicated.

The form of the kinetic energy functional is the most complicated part of the orbital-free theory. Finding precise kinetic energy functionals is the focus of much modern research~\cite{Dreizler1990, Engel2011, Wesolowski_2012, Karasiev_2014, Witt2018}. We opted for the simplest model that goes beyond the homogeneous electron gas case of the Thomas-Fermi method
\begin{equation}
    K[n] = \int_{\Omega_\mathrm{Sm}} \dd{\vb{r}}  C_\mathrm{TF}(\vb{r}) n ^{5/3}(\vb{r}) + \lambda_\mathrm{vW} \frac{\hbar^2}{8m^*} \frac{\norm{ \grad n( \vb{r} ) } ^2 }{n (\vb{r})} \,,
\end{equation}
where $C_\mathrm{TF}(\vb{r}) = (3 \pi^2)^{\frac{2}{3}} \frac{\hbar^2}{2m^*(\vb{r})} \frac{3}{5}$ is the Thomas-Fermi constant, $\lambda_\mathrm{vW}$ is the von Weizsäcker parameter and $m^*$ is the effective mass. The kinetic energy functional incorporates a TF term and the so-called von Weizsäcker correction~\cite{Geldart1986, VanZyl2013, Bartolotti2007, Witt2018, Low-dimensional2000, semiclassical}. This term captures the energy cost of rapid variation of the density in space and was originally introduced by von Weizsäcker to correct the TF method issues when applied to rapidly varying electron densities. 

In the literature, there has been much discussion on the value of the vW coefficient, $\lambda_\mathrm{vW}$, which works as a weight of the gradient-dependent vW term~\cite{VanZyl2013}. In the limit $\lambda_\mathrm{vW}\to0$, the TF method is recovered. In Ref.~\cite{Jones1971}, the response function of a uniform system of independent fermions is investigated, and it is shown that $\lambda_\mathrm{vW} = 1$ is exact in the limit of short-wavelength perturbations, whereas $\lambda_\mathrm{vW} = 1/9$ is exact in the limit of long-wavelength perturbations. Other analysis pointed to the value of $\lambda_\mathrm{vW}=1/5$ as the most adequate~\cite{Wang_PRB_1992}. 

In this application, we decided to treat $\lambda_\mathrm{vW}$ as a parameter of the model and empirically select a value in the $[0, 1]$ interval that agrees with SP simulations in simple geometries. The complete form of the energy functional and some of its extensions are further discussed in Ref.~\cite{semiclassical}.

% Not strictly a density functional theory

It is worth noting that a functional theory that takes this form is not, strictly speaking, a density functional theory, as the electric potential is explicitly included and cannot be easily transformed into a pure density functional theory. This is caused by the fact that metal parts (like gates or floating metallic islands) are present in nanoelectronic devices. These modify the boundary condition of the electric potential equation such that the PE Green's function becomes a complicated object generally not expressible in an analytic form. Including the electric potential explicitly avoids this problem. A more detailed discussion is included in Appendix \ref{sec:DFT}.

Moreover, when treating a closed system like an atom or a molecule, the minimization problem is characterized by the constraint
\begin{equation}
     \int_{\Omega} \dd{\vb{r}} n = N,
\end{equation}
where $N$ is the total number of electrons, that is fixed. The existence of a solution is not guaranteed when including an exchange and correlation potential~\cite{Lieb_1997}. We do not have such a constraint in our case. The total number of electrons $N$ is not fixed, but actually one of the results we want to determine. Moreover, it can also take fractional values.
% On the other hand, boundary conditions for the electrostatic potential are much more complicated due to the presence of metallic gates. 

To move from an optimization problem to a boundary value problem, we will use functional minimization. It is convenient to define the matter field $\psi = \sqrt{n}$ before proceeding. Note that $\psi$ is a real field defined as the square root of the density. It cannot be interpreted as a wavefunction as it does not carry information about the phases of the electrons. Therefore this theory cannot describe long-range interference effects, but it just includes a local correction to the electronic density due to quantum confinement effects.

The functional derivative $\fdv{E}{\varphi} = 0$ returns the Poisson equation while $\fdv{E}{\psi} = 0$ returns a partial differential equation that has the form of a nonlinear Schrödinger equation (NLSE). The system of coupled PDEs is then
\begin{widetext}
\begin{equation}
    \begin{cases}
- \div \varepsilon \grad \varphi = \rho_\mathrm{fx} - e \psi^2\,, \\
- \frac{\hbar^2 }{2}   \div \qty(   \frac{\lambda_\mathrm{vW}}{m^*} \grad \psi)  + \frac{5}{3} C_\mathrm{TF} \psi^{7/3} + \qty[- e \varphi +E_\mathrm{CBM}] \psi = 0\,.
    \end{cases}
\label{eq:PDEs}
\end{equation}
\end{widetext}

% Boundary conditions
The insulators are usually large gap semiconductors that are depleted for low enough bias voltages. Therefore, we can exclude insulator regions from the NLSE domain and assign the boundary condition of $\psi(\partial \Omega_\mathrm{I}) = 0$ to all semiconductor-insulator interfaces, where with the symbol $\partial \Omega_i$ we denote the surface of the region $\Omega_i$. We assume all metallic regions are described by ideal metals, and therefore the potential is homogeneous in these domains. When a voltage bias $V_i$ is applied to region $\Omega_{\mathrm{M}_i}$, a Dirichlet boundary condition can be used that takes the form 
$\varphi(\partial \Omega_{\mathrm{M}_i}) = \delta \mu_i + V_i$ where $\delta \mu_i$ is a material-specific constant modeling the Fermi energy difference between the metal in regions $\Omega_{\mathrm{M}_i}$ and the reference one. When $\Omega_{\mathrm{M}_i}$ is a floating island, the Neumann condition $\grad_{\bot} \varphi(\partial \Omega_{\mathrm{M}_i}) = 0$ fixes the electric field to be normal to the surface. Precise calibration of the band-offsets $\delta_i$ requires a careful comparison of numerical simulations and experimental results. Since this work is a proof of concept, the offsets $\delta \mu_i$ for metals are given arbitrary but reasonable values, and we refer to other references for the details of such procedure~\cite{Robertson2013}. The choice of material properties constants, including band offsets between semiconductors used to evaluate the conduction band minimum $E_\mathrm{CBM}$, are discussed in the Supplemental Material.

Metal-semiconductor interfaces are a vast topic that is still actively being researched~\cite{Monch2001}, and we have not found a convincing solution to the problem of including these interfaces in our method. We will first focus on the case where these interfaces are not present and assess the validity of the model while we postpone the discussion to Sec.~\ref{sec:metalBC}. 

\section{Calibration and benchmark}
% TF
Before considering the ETF theory, we analyze the limits of the TF method ($\lambda_\mathrm{vW}=0$) by simulating two simple but relevant geometries for 2DEG devices: a nanowire and a circular dot. For the 2DEG, we choose a semiconductor stack similar to the ones used in many modern experiments, e.g., Refs.~\cite{Poschl_PRB_2022a, Poschl_PRB_2022b, Banerjee2022}.  Fig.~\ref{fig:errors}(a) display the schematic of a cross-section of a 2DEG nanowire. Two \ch{Au} gates (yellow) serve the purpose of depleting the areas next to the \ch{Al} wire (dark gray). The two \ch{Au} gates and the \ch{Al} wire are separated by \ch{Hf O_2} dielectric (gray). Fig.~\ref{fig:errors}(b) shows the electron density in the system, simulated using a top gate voltage of \SI{-3}{\volt} with respect to the grounded \ch{Al} wire, while in Fig.~\ref{fig:errors}(c) we plot the TF error as defined in Eq.~\eqref{TF_error}. From this simulation, it is clear that the TF error does not satisfy $R_\mathrm{TF}\ll 1$, and thus the electron density is varying too quickly in space to justify the use of the Thomas-Fermi approximation. This behavior is consistent when trying different top gate voltages. In Figs.~\ref{fig:errors}(d), (e), and (f), we simulate a quantum dot on the same semiconductor stack. We apply a voltage of \SI{-0.25}{\volt} to the outer gate and \SI{0}{\volt} to the inner gate. In Fig.~\ref{fig:errors} e) and f), we show, respectively, the electron density in a plane located in the middle of the \ch{In As} layer and the TF error that approaches $1$ at the boundary to the depleted region.

% Calibration
\begin{figure*}[ht]
\centering
\includegraphics[width=1.0\textwidth]{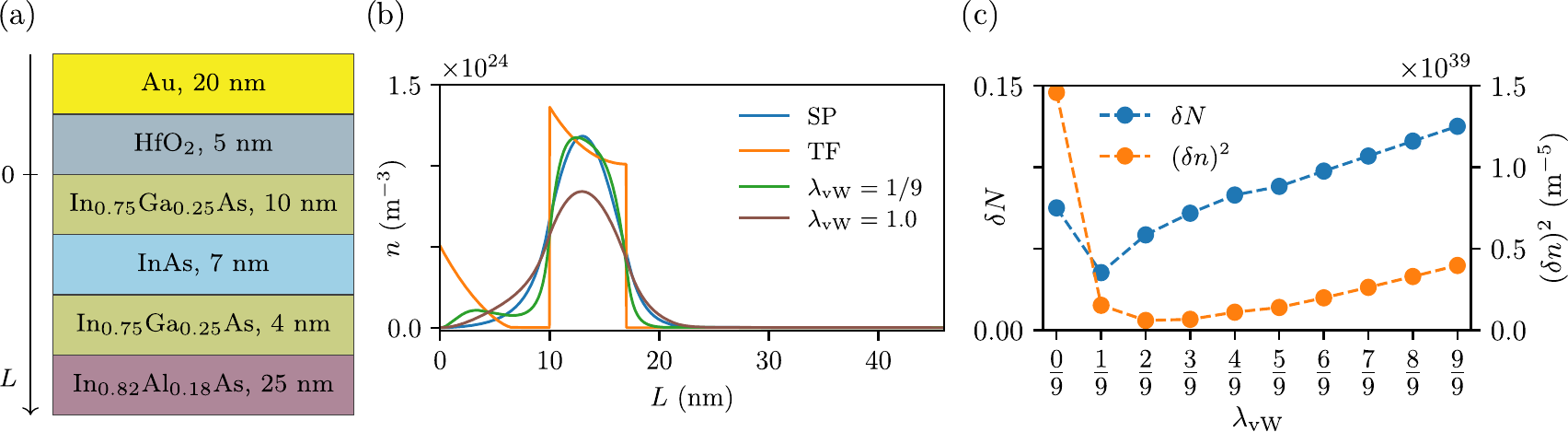}
\caption{Calibration of $\lambda_\mathrm{vW}$.
    (a) shows a schematic of the simulated semiconductor stack inspired by the one used in \cite{Banerjee2022} (not to scale). An \ch{Au} gate (yellow) is separated from the four semiconducting layers by a layer of oxide (gray). (b) shows the results of TF, ETF (with various $\lambda_\mathrm{vW}$), and SP simulations. (c) shows a comparison of different values of $\lambda_\mathrm{vW}$, using the two metrics defined in Eqs.~\eqref{eq_metric_2} and \eqref{eq_metric_3}. For all simulations the built-in bias at the interface is \SI{0.1}{\volt}.}
\label{fig:lambda_calibration}
\end{figure*}

After having assessed the need for a more elaborated kinetic energy functional than the Thomas-Fermi one, we now consider the ETF method. We start with the problem of determining the optimal value of $\lambda_\mathrm{vW}$ for the use case of electrostatic simulations of nanoelectronic devices. We considered a 2D translational invariant MOS device formed by a semiconductor heterostructure quantum well with an insulator layer and a metallic top-gate as shown in  Fig.~\ref{fig:lambda_calibration}(a). We study the electrostatic problem with the TF and ETF method with various $\lambda_\mathrm{vW}$ and compare the results with a simulation done with the self-consistent SP method. The density per unit area is shown in Fig.~\ref{fig:lambda_calibration}(b). 

To assess quantitatively the optimal value for the von Weizsäcker parameter, we introduce two metrics: the absolute difference of the density per unit area 
\begin{equation}
    \delta N \equiv \frac{\abs{N_\mathrm{vW} - N_\mathrm{SP}}}{N_\mathrm{SP}}\,,
\label{eq_metric_2}
\end{equation}
where $N = \int n\dd{x}$ is the total number of electrons per unit area, and the quantity
\begin{equation}
    (\delta n)^2 \equiv \int \qty( n_\mathrm{SP}- n_\mathrm{vW})^2\dd{x}\,,
\label{eq_metric_3}
\end{equation}
that takes into account the difference in the shape of the density profile. The results are shown in Fig.~\ref{fig:lambda_calibration}(c). In general, we find that low values of $\lambda_\mathrm{vW}$ in the interval $[0.05, 0.2]$ provide the best agreement with SP results. We decided to elect $\lambda_\mathrm{vW} = 1/9$ as the standard parameter because of the theoretical works backing this choice~\cite{Jones1971, VanZyl2013}.

\begin{figure}[hb]
\centering
\includegraphics[width=1.0\columnwidth]{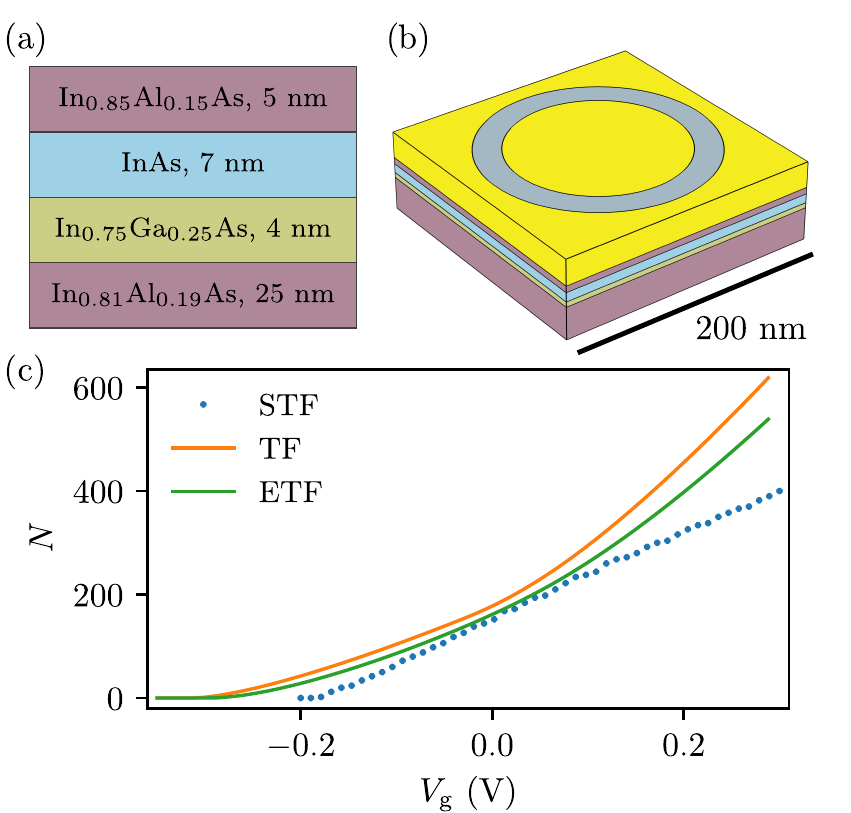}
\caption{Results from the dot-probe device simulation, using three different methods, i.e., TF, ETF, and STF. (a) shows the semiconducting stack used (not to scale), and (b) shows the geometry. (c) shows how electrons accumulate in the semiconducting stack as the inner gate voltage increases.}
\label{fig:dot}
\end{figure}

% Benchmark
We simulated a quantum dot shown in Fig.~\ref{fig:dot} with the TF, ETF, and STF methods to evaluate the accuracy of the method in a more realistic situation. Here we fix the outer gate to \SI{-0.35}{\volt} and consider the number of electrons in the dot as a function of the inner gate voltage. The results can be seen in Fig.~\ref{fig:dot}(c). Notice that the device under consideration does not show a dot-like behavior as the charge increases almost linearly with the gate voltage, as expected for a 2D system. This suggests that the lateral confinement induced by the gate system is not able to strongly confine the electrons. 

% The STF method used here consists of using the TF results for the electrostatic potential and then solving the Schrödinger equation once to get the electron density. % We see that all the methods predict depletion in the semiconducting stack at $\sim \SI{-0.35}{\volt}$ (the STF method predicts depletion until $\sim \SI{-0.2}{\volt}$), and as the inner gate voltage is increased, electrons start to accumulate in the stack.

Of the three methods, the TF method generally predicts the largest number of electrons, while the STF method predicts the lowest. The ETF results are intermediate between the two. Even though the ETF and STF methods do not overlap for all voltages, the two methods seem to have great compliance in the moderate filling regime.  

One important difference between the TF and ETF methods is that the TF method predicts a steep jump in the differential capacitance $C(V) = \dv{Q}{V}$ as the voltage increases. This happens as electrons at a particular voltage start accumulating in a previously classically forbidden region, while the ETF method, by allowing exponentially suppressed tails in the classically forbidden layers, prevents this from happening. This makes the ETF method less prone than the TF method to show unphysical behavior in semiconductor heterostructures. Thus the ETF method would, for instance, also be more appropriate to calculate capacitance compared to the TF method. In the setup considered here, we can thus conclude that the ETF method is superior to the TF method for simulating the number of electrons and calculating elements of the capacitance matrix.

\section{Semiconductor-metal boundary condition}
\label{sec:metalBC}
As is briefly described in the method presentation, the treatment of the interfaces between metals and semiconductors is an open problem~\cite{Monch2001}. Moreover, the electron density shows a strong dependence on the thickness of thin metallic films that afflicts heterostructures when placed in contact with clean metallic films~\cite{Mikkelsen2018}. Usually, the electrostatic problem of the charge distribution in the metal is not taken into account, and metallic parts are assumed neutral by assigning a Dirichlet boundary condition at the surface~\cite{Mikkelsen2018, Winkler2019}. We will consider the problem starting with a complete treatment of the metal-semiconductor interface and discuss the issues of this solution. Next, we will search for an approximate solution able to reproduce the important physical behavior of the electrostatics in the semiconductor.

\subsection{Complete treatment}
A complete treatment of the Sm-M interface can be formulated assuming that $\psi^2$ represents the conduction band electron density in both the semiconductor and the metal. We denote by $E_{F, \mathrm{M}}$ the bulk Fermi energy of the metal defined as $E_{F, \mathrm{M}} = \frac{\hbar^2 k_F^2}{2 m^*}$. We define $E_\mathrm{W}$ as the difference between the Fermi energy of the metal and the conduction band minimum of the semiconductor such that the local CBM takes the form
\begin{equation}
    E_\mathrm{CBM}(\vb r) = - E_\mathrm{w} \vb{1}_\mathrm{Sm}(\vb r) - E_{F, \mathrm{M}} \vb{1}_\mathrm{M}(\vb r)\,,
\end{equation}
where $\vb{1}_{\Omega_i}$ is the indicator function of region $\Omega_i$.

In addition, since in the metal the Fermi energy lies in the conduction band, the equilibrium bulk electron density needs to be compensated by a positive background that we assumed homogeneous and equal to 
\begin{equation}
    \rho_\mathrm{fx} = \rho_\mathrm{M} \vb{1}_M (\vb r),\qquad \rho_\mathrm{M} = \frac{e}{3 \pi^2} \qty(\frac{2m^* E_{F, \mathrm{M}}}{\hbar^2})^{3/2}\,,
\end{equation}
such that a homogeneous metallic system is neutral at equilibrium with an electron density equal to the one predicted by the Thomas-Fermi model. 
However, there are issues with this model when it is applied to a metal-semiconductor system. Since $\psi^2$ is the electron density in both materials, there will be a huge jump in its value at the interface, violating the assumption of slowly varying electron density. 

\subsection{Effective boundary treatment}
Since a complete model of the semiconductor-metal interface is not available and the boundary conditions for the ETF method are unknown, we explored the possibility of finding an effective way to treat the semiconductor-metal boundary.

% boundary conditions
A first approach consists in excluding the metal from the NLSE. In this way, $\psi^2$ models the electron density only in the semiconductor, and the problem reduces to finding an effective boundary condition by trial and error. We tested three options. If the interface is not  transparent, setting the density at the interface to zero $\psi(\partial\Omega_M)=0$ can be an acceptable boundary condition even in the metallic case. Alternatively, if we assume that the metal is not perturbed at all, we can impose that the density should continuously go to the unperturbed metal density at the interface $\psi(\partial\Omega_M) = \sqrt{n_M}$. Finally, we tried a Neumann boundary condition by fixing the change in electron density to zero at the semiconductor-metal interface $\partial \psi(\partial \Omega_M) = 0$.

%% lambda-sweep approach
These approaches cannot be applied if the metal has to be included, for example, because it is a floating part. In this case, we found that a mixed TF and ETF approach can be used, where we solve for the electron density in both the semiconductor and the metal, but we allow $\lambda_\mathrm{vW}$ to vary in space. The idea of promoting $\lambda_\mathrm{vW}$ to an inhomogeneous field has already been considered in more traditional application fields of OF density functional theory~\cite{Plumer_JPC_1983, Plumer_JPC_1985}.

Since metals have an extremely large electron density compared to that of the semiconductor, there will be an extremely large gradient of the electron density at the very interface. Since the vW-correction of the energy functional is proportional to the absolute value of this gradient squared, the vW-term will be extremely large here and thus essentially penalize the energy functional, trying to remove the abrupt change in electron density. However, this abrupt change in electron density at the Sm-M interface is what we would expect of the system and should thus not be removed. A solution to this could be to change $\lambda_\mathrm{vW}$ through the stack to diminish the correction where we expect large gradients. We dubbed this method $\lambda_\mathrm{vW}$-sweep. In the metal, we expect an extremely large electron density (when compared to the semiconductor) that is only weakly perturbed by being in contact with a semiconductor. Changes are thus slow in space on the length scale of the Fermi wavelength, and the electron density in the metal can be effectively described by the TF method, i.e., assigning $\lambda_\mathrm{vW} \simeq 0$ in the metal while using a finite $\lambda_\mathrm{vW}$ in the semiconductor. This circumvents the problem of assigning a boundary condition to the interface. In the bulk of the semiconductor, we use $\lambda_\mathrm{vW} = 1/9$, and in the aluminum we set $\lambda_\mathrm{vW}$ to zero (for convergence reasons, we use a small but non-zero $\lambda_\mathrm{vW}$ of $2 \cdot 10^{-8}$). 

% TEST MODEL
\begin{figure}[hb]
\centering
\includegraphics[width=1.0\columnwidth]{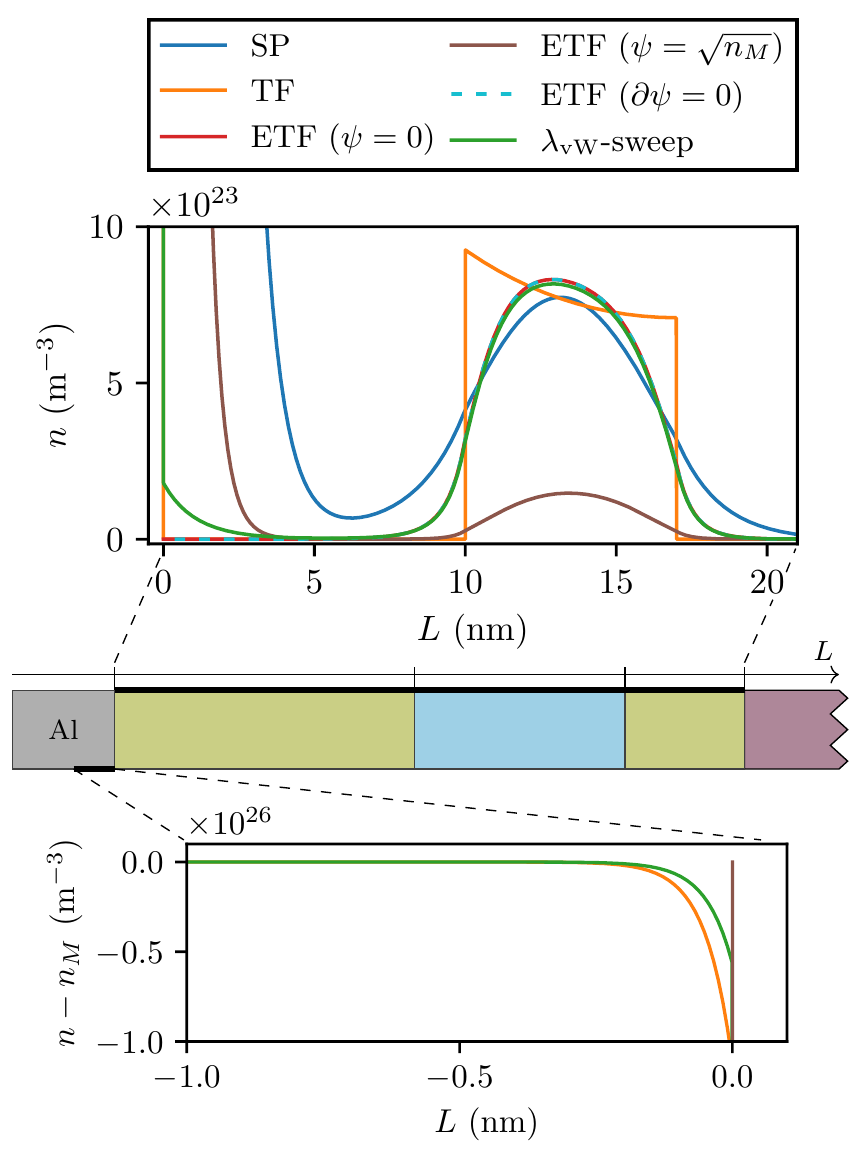}
\caption{Simulation of the semiconducting stack in Fig.~\ref{fig:lambda_calibration} (a), with a \SI{5}{\nm} layer of aluminum on top. The upper plot shows the electron density in the semiconducting stack, while the lower one shows the electron density in the metal. 6 different ways of treating the semiconductor-metal interface are simulated: Three ETF simulations with different boundary conditions at the interface ($\psi = 0$, $\psi = \sqrt{n_\mathrm{Al}}$, and $\partial \psi = 0$), one TF simulation, one SP simulation that is solved only in the semiconductor, and one ETF simulation where the value of $\lambda_\mathrm{vW}$ is swept such that it is $1/9$ in the semiconductor and $0$ in the metal.
}
\label{fig:metal-sm}
\end{figure}

To test the method above, we considered a 2DEG system built with the semiconducting stack from Fig.~\ref{fig:lambda_calibration} with a \SI{5}{\nano\meter} layer of \ch{Al} deposited on top as shown in Fig.~\ref{fig:metal-sm}. In the cases where the metal is included in the density part, we set the Fermi energy to $E_{F, \ch{Al}}=\SI{11.6}{\electronvolt}$ and neutralized the bulk charge with a positive background. With this value, the bulk electron density of \ch{Al} is roughly $n_{M} = \SI{1.8e29}{\meter^{-3}}$. The electron density simulated for all the cases described can be seen in Fig.~\ref{fig:metal-sm}. For comparison, we included the result of a SP simulation. 
% We set a gate voltage \SI{1}{\nano\meter} to the right of the semiconductor-metal interface to avoid taking into account the surface dipole.
% result that is obtained however we fixed the electron density to zero at the semiconductor-metal interface to avoid the problem of resonances that is present in semiconductors in contact with clean thin metallic films~\cite{Mikkelsen2018}.

% CONCLUSIONS OF SECTION
In the top plot of Fig.~\ref{fig:metal-sm}, we see how the results for the $\psi = 0$ and $\partial \psi = 0$ boundary conditions are similar to the ones obtained with $\lambda_\mathrm{vW}$-sweep model, and how all of these methods seem to provide a good agreement with the SP method. The Dirichlet boundary condition $\psi = \sqrt{n_\mathrm{Al}}$ approach seems to give an erroneously small electron density in the well. This is caused by the extreme change in electron density that has to be neutralized entirely in the semiconductor. This causes an excess of negative charge near the interface that is left unbalanced. The $\lambda_\mathrm{vW}$-sweep method makes it possible to avoid this problem since the electron density in the metal is now described by the TF method and, thus, no boundary condition is required at the semiconductor-metal interface. As can be seen, even if the depletion in the metal is localized in a tiny region at the interface, given the extremely high bulk density, the decrease causes a strong dipole at the interface that strongly changes the electrostatics restoring a result closer to the SP simulation. We can conclude that if the semiconductor layer closer to the metal is classically forbidden, both $\psi(\partial \Omega_M)=0$ and $\partial\psi(\partial \Omega_M)=0$ are acceptable effective boundary conditions for the orbital-free theory. If the metal has to be explicitly included, a $\lambda_\mathrm{vW}$-sweep can be used for this purpose. We ruled out the  $\psi(\partial \Omega_M)=\sqrt{n_M}$ option as it does not provide comparable results.

\section{General consideration on computational complexity}
Before concluding, we briefly discuss the convergence properties of ETF methods, and in general, OF methods, in comparison to TF and SP. Estimating the space and time complexity of these methods can be complicated, yet by assuming that the electric potential is discretized using $N_e$ degrees of freedom and the charge distribution on $N_q$ degrees of freedom, we can make some general observations.

The TF method is represented using only the electric potential, leading to a space complexity of $O(N_e)$. The method requires solving a relatively simple nonlinear PDE for the electric potential, which generally requires iterative methods to converge, making it difficult to estimate the convergence rate.

In contrast, the ETF method is represented as a functional of both the electrostatic field and the matter field, leading to a space complexity of $O(N_e + N_q)$. The coupled PDEs that need to be solved are more complex than the one in the TF method. In some cases, it is possible to decouple the two PDEs and solve them separately and iteratively (using a so-called segregated or partitioned approach in contrast to a fully coupled/monolithic one) in order to optimize the convergence speed.

We assume that only $k$ electrons are considered for the SP method, leading to a space complexity of $O(N_e + k N_q)$. The calculation is dominated by the solution of the Schrödinger equation, and assuming the use of an iterative solver like Arnoldi iteration, the time complexity is bounded from below by $O(kN_q^2)$. The cost of diagonalization is needed for each step of the iteration, together with the solution of the Poisson equation, which generally has a negligible cost compared to diagonalization. The solution strategy is inherently segregated, and this can cause potential instability problems that need to be addressed with special care~\cite{Armagnat_2019}. It is important to note that in the case of large systems, not only the number of degrees of freedom increase, but, in general, it is also necessary to increase the number of states $k$ considered.

In summary, it is clear that the computational cost of OF methods increases more slowly with the size of the system compared to the SP method, making it a more attractive option for large-scale simulations. Moreover, the expression in variational form is amenable to the application of many convergence optimization techniques commonly used for general PDEs that are not easily applicable to diagonalization problems like multigrid methods, preconditioning, and in particular, domain decomposition.

\section{Conclusions}
In this work, we investigated orbital-free methods for the solution of the electrostatic problem for nanoelectronic devices. We checked that the widespread Thomas-Fermi method, regarded as the simplest orbital-free method, is often applied outside its range of validity. The most notable case of the approximation breakdown occurs at the interfaces with classically forbidden regions, like insulators, where the density of electrons has abrupt jumps. 

To achieve a more accurate description of the density profile in these cases, we considered the extended Thomas-Fermi (ETF) method that includes the von Weizsäcker correction of the kinetic energy functional. This correction can significantly increase the precision of the density profile near interfaces. 

We addressed the question of the optimal value of the $\lambda_\mathrm{vW}$ parameter by studying a simple 2DEG system and found that the theoretically motivated value of $\lambda_\mathrm{vW}=1/9$ provides a good agreement also in the practical example cases considered. By applying the method to the simulation of realistic device geometries, we found that the ETF method provides density profiles closer to the ones calculated with the Schrödinger-Poisson method than the density profile provided by the TF method. 

SP methods can be computationally expensive as they require explicit diagonalization of the Hamiltonian, whereas the predictive power of the TF method is poor due to the perfect local behavior of the energy functional. Therefore, the ETF method represents a good compromise in terms of computational speed and predictive power. orbital-free methods are an often neglected alternative for electrostatic simulations of nanoelectronic devices, which are useful to handle large systems since the problem can be nicely expressed in variational form and implemented on any finite element solver.

Other effects like strong spin-orbit coupling~\cite{Brack_PR_1985}, finite temperature~\cite{Brack_PRL_1984, Bartel_NPA_1985}, non-parabolicity~\cite{Zollner1986, Altschul1992} and exchange interaction~\cite{Ubensee1988}, can be included by using more advanced energy functionals.

\section{Acknowledgments}
W. S. and A. M. acknowledge Georg Winkler for his help and support through this work, and Andreas Pöschl, Alisa Danilenko, and Andrea Giuseppe Landella for useful discussion. The authors acknowledge Microsoft Quantum for support and computational resources. This work. was supported by the Danish National Research Foundation, the Danish Council for Independent Research \textbar  Natural Sciences.

\appendix
\section{General orbital-free functional theory}
\label{sec:DFT}
As mentioned in the main text, the theory proposed is not strictly a density functional theory as the electric potential appears as an argument of the energy functional, differently from what is common in electronic structure calculations. In principle, it is still possible to eliminate the electric potential from the formulation, but the presence of metallic gates makes the application of the procedure extremely complicated and cumbersome. 

To illustrate the procedure, we can start with the generic orbital-free functional provided by Eq.~\eqref{eq:energy_functional} and proceed to find the minimizing condition for $\varphi$. We get
\begin{equation}
    \frac{\delta E}{\delta \varphi} = \rho (\vb{r}) + \grad \cdot ( \varepsilon \grad \varphi (\vb{r})) = 0\,,
\end{equation}
which is exactly the PE. This is abstractedly solved by the fundamental solution $G(\vb r)$ such that
\begin{equation}
     - \grad ^2 G(\vb{r}, \vb{r}') = \delta (\vb{r}'),
\end{equation}
where $\delta (\vb{r}')$ is the Dirac's delta function.

Therefore, we can express the electric potential as 
\begin{equation}
    \varphi (\vb{r}) = \int G(\vb{r},\vb{r}') \rho(\vb{r})\rho(\vb{r}') \dd{\vb{r}'}
\end{equation}
and consequently rewrite the energy functional as a function of the charge density alone 
\begin{equation}
\begin{split}
    E[ \rho ] = &K[\rho] + \int V_\mathrm{ext} \rho (\vb{r}) d \vb{r} \\
    + &\int \rho(\vb{r}) G (\vb{r}, \vb{r}')\rho(\vb{r}') \dd{\vb{r}}\dd{\vb{r}'}\,.
\end{split}
\end{equation}

This density functional theory formulation is exactly equivalent to the orbital-free theory that is the object of this paper. However, it relies on the fundamental solution $G(\vb{r}, \vb{r}')$ that is an extremely complicated object in any real case since in almost any case inhomogeneous permittivity and metallic gates with nontrivial shapes are present in the system. 

For electronic structure calculations, this is not the case and indeed the PE equations have the simple fundamental solution 
\begin{equation}
    G(\vb{r}, \vb{r}') = \frac{1}{4 \pi \varepsilon \vert \vb{r} -\vb{r'} \vert}\,.
    \label{new_correction_5}
\end{equation}
With this simplification, we can write the total energy functional as a functional of $\rho$ only 
\begin{equation}
\begin{split}
    E[ \rho ] = E_\mathrm{k} [\rho] + \int V_\mathrm{ext} \rho (\vb{r}) d \vb{r} \\
    +  \int \frac{\rho (\vb{r}) \rho (\vb{r}')}{4 \pi \varepsilon } \frac{1}{\vert \vb{r} - \vb{r}' \vert} d \vb{r}'\,,
\end{split}
\end{equation}
as we have seen that the functional minimization with respect to $\varphi$ directly gives Poisson's equation. This means that we can now minimize the functional with respect to $\rho$, and couple this with Poisson's equation to capture both the physical behavior from $\varphi$ and $\rho$. Thus we have decoupled the problem and can now write the functional with respect to $\rho$ only.

This is exactly the form that we implicitly assume in the calculations above and, thus, the functional we use to derive the TF and ETF methods. This result is exact in free space, as it is under the assumption of $\varepsilon$ being constant and under the assumption of no boundary conditions. However, this is not correct in any real device. 

\bibliography{bibliography.bib}
\clearpage


\section*{Supplementary material}

# Supplemental Material to "Orbital-free approach for large-scale electrostatic simulations of quantum nanoelectronics devices"

Waldemar Svejstrup,[1] Andrea Maiani,[1] Kevin Van Hoogdalem,[2] and Karsten Flensberg[1]

[1]*Center for Quantum Devices, Niels Bohr Institute,
University of Copenhagen, Copenhagen, Denmark*
[2]*Microsoft Quantum Lab Delft, Delft University of Technology, Delft, Netherlands*

(Dated: February 2023)

## DETAILS OF THE SIMULATIONS

In this section of the Supplemental Material, we provide details on the numerical details of the simulations presented in the paper.

### A. Details of the model

For the simulations in Fig. 1 and Fig. 2, we have assumed alignment between the Fermi level of the gate and the upper barrier ($In_{0.75}Ga_{0.25}As$) and fixed this to zero energy. The conduction band minimums of the other materials are thus in relation to this zero-point. This means that we have introduced an effective conduction band minimum, $E_{CBM}$, that is in relation to this zero-point. Therefore, the effective conduction band minimum of the $In_{0.75}Ga_{0.25}As$ is the reference level, as can be seen in Tab. I. In Tab. I all other material properties used in the simulations are also found. For the simulations in Fig. 3, we will be using a different upper barrier ($In_{0.85}Al_{0.15}As$) and give this an off-set of $-0.0325$ eV with respect to the gate. This is also summarized in Tab. I.

|  | $\varepsilon/\varepsilon_0$ | $m_{eff}/m_e$ | $E_{CBM}$ [eV] |
|---|---|---|---|
| $In_{0.75}Ga_{0.25}As$ | 14.76 | 0.035 | 0.00 (ref) |
| InAs | 15.15 | 0.026 | -0.205 |
| $In_{0.85}Al_{0.15}As$ | 14.39 | 0.038 | 0.079 |
| $In_{0.82}Al_{0.18}As$ | 14.23 | 0.041 | 0.136 |
| $In_{0.81}Al_{0.19}As$ | 14.18 | 0.042 | 0.155 |
| $HfO_2$ | 25.00 | 1.000 | 4.793 |

TABLE I. Relevant material parameters used in the simulations. For simulations using $In_{0.75}Ga_{0.25}As$ as upper barrier we have $\delta\mu_i = 0$ eV, and for simulations using $In_{0.85}Al_{0.15}As$ as upper barrier we have $\delta\mu_i = -0.1113$ eV. The numerical values are taken from various sources [1–5].

For Fig. 1 (a), (b), and (c), all mesh elements are $0.5$ nm, the two gates are fixed at $-3$ V, and the aluminum wire is grounded. The aluminum wire is $130$ nm wide. We set the relative tolerance of the solver to $10^{-3}$.

For Fig. 1 (d), (e), and (f), all mesh elements in the oxide and gates are $5$ nm, in the semiconductors the mesh elements are $5$ nm in x and y-directions, and $1$ nm in the z-direction. Outer gate fixed at $-0.25$ V and inner gate at $0$ V. The inner gate radius is $65$ nm, and the oxide ring radius is $135$ nm. We set the relative tolerance of the solver to $10^{-3}$.

For Fig. 2, all mesh elements are of length $0.002$ nm. Gate fixed at $0.1$ V. The relative tolerance of the solver used for the TF and ETF simulations is set to $10^{-3}$ while the relative tolerance of the SP solver is set to $10^{-5}$.

For Fig. 3, the TF and ETF simulations have mesh elements in the oxide and gates of $5$ nm, while in the semiconductor regions, the mesh elements are $5$ nm in the x and y direction, and $1$ nm in the z-direction. The STF simulations have a mesh of $5$ nm in the gates, $3$ nm in the oxide, and $2$ nm in the semiconductors for the first TF simulation. Afterward, the Schrödinger equation is solved using a meshing of $3$ nm in x- and y-directions and $1$ nm in the z-direction, in the semiconductor. For the ETF simulations, $\lambda_{vW} = 1/9$ is used. Outer gate fixed at $-0.35$ V. The radius of the inner gate is $65$ nm, and the radius of the ring of oxide is $85$ nm. We set the relative tolerance of the solver used for the TF and ETF simulations to $10^{-4}$ while we used a relative tolerance of $10^{-9}$ for the solver used for the TF part of the STF simulation.

For Fig. 4, the TF and ETF simulations have mesh elements of length $0.001$ nm in the semiconductor and $0.00002$ nm in the gate. The SP simulation has mesh elements of length $0.01$ nm in both the semiconductor and the gate. Gate voltage fixed at $0$ V. $\lambda_{vW} = 1/9$ is used for the ETF simulations in the semiconductor, and $\lambda_{vW} = 2^{-8}$ is used for the

$\lambda_{\text{vW}}$-sweep in the metal (for numerical reasons we chose a very small value for $\lambda_{\text{vW}}$ in the metal, instead of completely fixing it to zero). Relative tolerance of the solver used for these simulations: $10^{-5}$.

### B. Details of orbital-free simulations

All TF and ETF simulations in this paper are made with COMSOL Multiphysics. For the electrostatic simulations, we have used the electrostatic module, and, to include the functional minimization in the ETF method, we have used the Weak Form PDE module. For the meshing in the 3D cases, we have used a mix of free tetrahedral and swept meshing. We have used COMSOL's built-in meshing algorithms to construct a free tetrahedral mesh in every region with gates since these have non-trivial shapes. We then swept this free tetrahedral mesh through the semiconducting stack using the swept mesh option of COMSOL. For the 2D case, we have meshed everything using the free triangular option. For the 1D cases, we have simply divided the system into a set of equally separated lattice points.

### C. Details of SP simulations

All SP and STF simulations are performed using the FENICS package [6] for finite element modeling of the electrostatics and Kwant for solving the Schrödinger equation [7].


\end{document}